\documentclass[pra, preprint,aps]{revtex4-1}

\usepackage{amsmath}
\usepackage{amsfonts, mathrsfs}
\usepackage{amssymb}
\newcommand{\nlse}{nonlinear Schr\"odinger equation }
\renewcommand{\j}{\psi(x,t)}
\newcommand{\p}{\pi(x,t)}
\renewcommand{\d}{\partial}
\newcommand{\am}{a_{-}}
\renewcommand{\ap}{a_{+}}
\newcommand{\Pp}{\mathscr{P}}

\begin{document}
\title{Quantum theory of nonlocal \nlse}

\author{Vivek M. Vyas}
\email{vivekmv@imsc.res.in}
\affiliation{Institute of Mathematical Sciences, CIT Campus, Taramani, Chennai 600 113, INDIA}

\author{Zodinmawia}
\email{zodin@imsc.res.in}
\affiliation{Institute of Mathematical Sciences, CIT Campus, Taramani, Chennai 600 113, INDIA}

\date{\today}

\begin{abstract}
Nonlocal nonlinear Schr\"odinger model is quantised and exactly solved using the canonical framework. It is found that the usual canonical quantisation of the model leads to a theory with pathological inner product. This problem is resolved by constructing another inner product over the vector space of the theory. The resultant theory is found to be identical to that of nonrelativistic bosons with delta function interaction potential, devoid of any nonlocality. The exact eigenstates are found using the Bethe ansatz technique.     
\end{abstract}

\maketitle

\section{Introduction} 
The study of exactly soluble and integrable models in both classical and quantum mechanics has been a subject of interest for long. A primary reason behind this interest, even though such models may or may not always be experimentally realisable, is the fact that it is possible to construct exact solutions to the model, which provides significant insight into physics of the model. The nonlinear Schr\"odinger equation, which reads:
\begin{equation}
  \left( i \frac{\d}{\d t} + \frac{\d^2}{\d x^2} \right) \j + g {\psi}^{\ast}(x,t) \j^2 = 0,  
\end{equation} 
is one such well studied integrable model, in both classical as well as quantum framework. 
Recently, a variant of this model, called nonlocal nonlinear Schr\"odinger equation was proposed by Ablowitz and Mussliman \cite{ablo}, and its classical dynamics was studied. It was shown, using the inverse scattering method, that akin to nonlinear Schr\"odinger equation, this model is also integrable. Further, exact solutions to this model were also constructed. It was also shown that this model, like many other integrable models, is a member amongst a hierarchy of integrable models which are called nonlocal nonlinear Schr\"odinger hierarchy.

The aim of this paper is to construct and study this model in the quantum framework. In particular, the nonlocal nonlinear Schr\"odinger equation shall be interpreted as a field equation, with the field being a linear operator on a Hilbert space. It is found that, the usual canonical route of quantisation for this model, leads to a theory with pathological inner product. This problem is rectified by constructing another inner product over the vector space. The quantum theory thus obtained is found to be identical to that of nonrelativistic bosons with delta function interaction. This identification leads one to conclude that the model is devoid of nonlocality. The exact eigenstates of the model are constructed using the Bethe ansatz technique.

\section{Classical theory}

The nonlocal \nlse for a complex field $\j$ is given by:
\begin{equation} \label{nnlse}
  \left( i \frac{\d}{\d t} + \frac{\d^2}{\d x^2} \right) \j = 2 c {\psi}^{\ast}(-x,t) \j^2.
\end{equation}
The theory is said to be nonlocal, due to the nonlinear term, which depends on both $\j$ and ${\psi}^{\ast}(-x,t)$. This equation can be derived as an Euler-Lagrange equation from the action \cite{bogol}:
\begin{align}
  \mathrm{S} &= \int dt dx \: \mathscr{L}, \\ 
\mathscr{L} &=  {\psi}^{\ast}(-x,t) \left( i \frac{\d \j}{\d t} + \frac{\d^2 \j}{\d x^2} \right) - c \left( {\psi}^{\ast}(-x,t) \j \right)^{2}.
\end{align}
Canonically conjugate momentum corresponding to $\j$ is given by $\p$\footnote{In this manuscript throughout, subscript stands for partial differentiation with respect to the variable, for example $f_{x}(x,t) = \frac{\d f(x,t)}{\d x}$.}:
\begin{equation}
  \p = \frac{\d \mathscr{L}}{\d \psi_{t}(x,t)} = i {\psi}^{\ast}(-x,t). 
\end{equation}
This allows one to construct the Hamiltonian for the theory, through the Legendre transform from the above action \cite{bogol}:
\begin{align}
\mathrm{H} &= \int^{\infty}_{-\infty} dx \: \mathscr{H}, \\
&= \int^{\infty}_{-\infty} dx \: \left( - i \pi_{x}(x,t) \psi_{x}(x,t) - c (\p \j)^{2} \right).
\end{align}
The fundamental equal time Poisson brackets for the dynamical fields $\j$ and $\p$ are given by:
\begin{align}
&  \{ \j, \pi(y,t) \} = \delta(x-y), \\
&  \{ \j, \psi(y,t) \} = 0, \quad \{ \pi(x,t), \pi(y,t) \} =0. 
\end{align}
It can be checked that (\ref{nnlse}) indeed can be obtained as an evolution equation for $\j$: $\psi_{t}(x,t) = \{ \j, H \}$, using the above Hamiltonian.
As mentioned earlier, it was shown by Ablowitz \emph{et. al.} \cite{ablo}, using inverse scattering theory, that this theory is integrable (in the sense of Liouville). This means that the theory admits an infinite set of conserved quantities (such that their mutual Poisson brackets are vanishing), first few of which are as below:
\begin{align}
  N &= -i \int^{\infty}_{-\infty} dx \: \j \p, \\
  P &= \int^{\infty}_{-\infty} dx \: \pi(x,t) \psi_{x}(x,t), \\ 
  H &=\int^{\infty}_{-\infty} dx \: \left( - i \pi_{x}(x,t) \psi_{x}(x,t) - c (\p \j)^{2} \right).
\end{align}
These conserved quantities generate transformations on the dynamical fields $\j$ and $\p$, and hence on any function of these dynamical fields. The infinitesimal transformation generated by a conserved quantity $Q$ (also called the conserved charge) on some dynamical variable $O$ is defined by $\delta_{Q} O(\j,\p) = \epsilon \{ O, Q \}$, where $\epsilon$ is a real infinitesimal. It can be checked that the transformations generated, by the above mentioned charges are such that, the action (in the first order form):
\begin{equation}
  S = \int dx dt \: \left( \p \psi_{t}(x,t) - \mathscr{H} \right),
\end{equation} 
remains invariant: $\delta S = 0$ \cite{schwing}. It is possible to give a converse argument for the existence of the conserved charges. It can be stated that the action of the theory is such that, it remains invariant under continuous transformations, as defined above. The celebrated Noether theorem then states there must exist corresponding conserved charges \cite{bogol}, and it can be easily seen that these are indeed the ones mentioned above. 

The transformations generated by $N$ are phase transformations: $\delta_{N} \j = - i \epsilon \j$ and $\delta_{N} \p =  i \epsilon \p$, and hence it can be said that, the invariance of action of the theory under these phase transformations implies existence of $N$ as a conserved charge. In a first quantised framework, existence of such a conserved charge in a different model was also noticed in Ref. \cite{abhi}. 

Similarly, existence of conserved field momentum $P$, clearly implies that the theory is invariant under spatial translations $x \rightarrow x + \text{constant}$, since $P$ generates spatial translations: $\delta_{P} \j = \epsilon \{ \j, P \} = \epsilon \psi_{x}(x,t)$. The role of $P$ is analogous to that of $H$ which generates temporal translations. This result is particularly important since, spatial translational invariance of the theory is difficult to immediately see from equation (\ref{nnlse}) itself.

\section{Quantum theory}

Construction of quantum theory following the canonical route is straightforward, and begins by postulating existence of field operators on an assumed Hilbert space, such that following canonical commutation relations hold:
\begin{align}
&  [ \j, \pi(y,t) ] = i \delta(x-y), \\
&  [ \j, \psi(y,t) ] = 0, \quad [ \pi(x,t), \pi(y,t) ] =0. 
\end{align}
Here, as argued by Dirac \cite{dir}, $\pi(x,t)$ is given by $i {\psi}^{h}(-x,t)$, where ${\psi}^{h}$ stands for Hermitean conjugate of $\psi$. These commutation relations imply that, the quantum theory is going to be a theory of interacting bosons. One can similarly go ahead and quantise the theory using anticommutators, to have a theory of interacting fermions. However, in this paper, the discussion will only be confined to the bosonic case. It must be noted that, corresponding to the conserved charges found in the classical case, there exists appropriate (normal ordered) operators in the quantum case.
Let the Fourier transforms of field operators (at $t=0$) be given by:
\begin{align}
  \psi(x,0) &= \int \frac{dp}{2 \pi} e^{i px} a_{-}(p), \\
  \pi(x,0) &= i\int \frac{dp}{2 \pi} e^{- i px} a_{+}(p).
\end{align}
It must be pointed out that, in the case of nonlinear Schr\"odinger equation, since $-i\p$ and $\j$ are adjoint of each other, it turns out that $\ap$ and $\am$ get related by the adjoint operation: $\ap^{h}(p) = \am(p)$. However, in this case, it can be easily seen that, these operators get related in a peculiar way under adjoint operation: $\ap^{h}(p) = \am(-p)$.
This leads to pathology in the construction of quantum theory, as it will be soon apparent. Working with operators $\ap(k)$ and $\am(k)$ makes the quantum treatment more transparent, since these obey following algebra:
\begin{equation}
\begin{aligned}
\label{fock}
  &[ N, \am(p) ] = - \am(p), \\ &[N, \ap(p)] = \ap(p), \\
  &[ \am(p), \ap(q) ] = 2 \pi \delta(p-q),
       \end{aligned}
\end{equation}
where $N = \int \frac{dp}{2 \pi} \ap(p) \am(p)$. Owing to this algebra, one can now interpret conserved charge $N$ as the number operator, whereas $\ap$ as the particle creation and $\am$ as particle annihilation operators. Hamiltonian can now be expressed as:
\begin{equation}
\begin{aligned} 
\label{Hamil}
  H &= H_{0} + V, \\
  H_{0} &= \int \frac{dp}{2 \pi} w_{p} \ap(p) \am(p), \\
  V &= 2 \pi c \int \frac{d p_{1}}{2 \pi} \frac{d p_{2}}{2 \pi} \frac{d p_{3}}{2 \pi} \frac{d p_{4}}{2 \pi} \: \delta(p_{1}+p_{2}-p_{3}-p_{4}) \ap(p_1) \ap(p_2) \am(p_3) \am(p_4).
\end{aligned}  
\end{equation}
Interestingly, since $[N, H_{0}]=0$, the eigenstates for number operator $N$ as well as that of the free Hamiltonian $H_{0}$ can be easily constructed, using algebra (\ref{fock}) by repeated application of creation operator $\ap(k)$ on the vacuum state $|vac \rangle$. Further, since $N$ and $H$ commute, it implies that using the states of Fock space (the eigenspace of $H_{0}$ and $N$), it would be possible to construct eigenstates of $H$. One simple way of realising this, is by using the Lippmann-Schwinger equation \cite{sakurai} to construct the ``in'' eigenstates of the total Hamiltonian $H$ from the knowledge of $n$-particle Fock state:
\begin{equation} \label{sol}
  | \Psi( k_{1}, \cdots, k_{n}) \rangle = \sum^{\infty}_{l=0} \left( G_{0}(w_{k}) V \right)^{l} |k_{1}, \cdots, k_{n} \rangle, 
\end{equation}
where $G_{0}$ is the free particle Green's function operator:
\begin{equation}
  G_{0}(w) = \frac{1}{w - H_{0} + i \epsilon},
\end{equation}
and $|k_{1}, \cdots, k_{n} \rangle = \ap(k_{1}) \cdots \ap(k_{n}) |vac \rangle$ is $n$-particle Fock state with energy $w_{k} = (k^{2}_{1} + \cdots +k^{2}_{n})$ and momentum $(k_{1} + \cdots + k_{n})$. 

In this approach, inorder to construct the eigenstates of $H$, the knowledge of the Fock space is essential. As mentioned earlier, the Fock space can be constructed from the vacuum state by repeated application of creation operator. However, since creation and annihilation operators are connected by adjoint operation in a peculiar manner, it turns out that the Fock space has a pathology, for all the states apart from vacuum. This can be clearly seen by looking at single particle sector, which is obtained by action of creation operator $\ap(p)$ on vacuum $|vac\rangle$. Vacuum is assumed to be a unit normalised state such that: $\am(p) |vac \rangle = 0$. The inner product between two single particle states is given by:
\begin{align} \nonumber
  \langle p | q \rangle = &\langle vac | \ap^{h}(p) \ap(q) |vac \rangle \\ \nonumber
   = & \langle vac | \am(-p) \ap(q) |vac \rangle \\    \nonumber
   = & 2 \pi \delta(p+q).
\end{align}
Thus the norm of a single particle state $|p\rangle$ is given by: $\langle p | p \rangle = 2 \pi \delta(2p)$. Note that state $|p\rangle$ is a momentum eigenstate with nonvanishing eigenvalue $p$. So one finds that, all the single particle states are zero normed:  $\langle p | p \rangle = 0$, for all $p \neq 0$. However, by definition, a zero normed vector is the one, which is orthogonal to all the vectors including itself. Note that although state $| p \rangle $ is zero normed, it is not orthogonal to all the vectors, since $\langle p| -p \rangle \neq 0$. The same argument can be easily extended to multiparticle states as well. This essentially means that the inner product defined over the Fock space, following Dirac \cite{dir}, is pathological. 

\subsection*{Redefinition of inner product}

Above one saw that the Fock space of the theory, has a pathological inner product. 
Inorder to cure the issue, and salvage the quantum theory, the approach originally due to Heisenberg \cite{hei, barton, lee} is followed in this paper. It rests on the idea that, problems related to normalisation can be resolved by working with a different norm. This new inner product is defined as: $\langle A | B \rangle_{O} = \langle A |O| B \rangle$, where $O$ is a suitable operator, chosen in such a manner that the vector space becomes a Hilbert space with respect to this inner product. It turns out that this approach can be successfully used in the present case as well.

Let $\Pp$-inner product between two states be defined as: $\langle A | B \rangle_{p} = \langle A |\Pp | B \rangle$, where the operator $\Pp$ is the parity operator. Parity operator $\Pp$ is a unitary operator, defined such that:
\begin{align} \nonumber
  \Pp \psi(x,t) = \psi(-x,t) \Pp \quad \text{and} \quad \Pp \pi(x,t) = \pi(-x,t) \Pp.
\end{align}
Further, $\Pp |vac \rangle = 0$ and $\Pp^{-1} = \Pp$. From above it follows that:
$\Pp \ap(p) = \ap(-p) \Pp$ and $\Pp \am(p) = \am(-p)\Pp$.

With this definition, it is straightforward to see that:
\begin{align} \nonumber
 \langle p | q \rangle_{p} &= \langle p | \Pp | q \rangle \\ \nonumber
 & = \langle p | - q \rangle \\
 & = 2 \pi \delta(p-q). 
\end{align}
This shows that the single particle states are positive normed states with respect to $\Pp$-inner product, and form an orthonormal set. Thus the problem of zero-normed states and orthogonality stands resolved.

Since the definition of adjoint of an operator is with respect to a given inner product, working with $\Pp$- inner product leads to identification of corresponding $\Pp$-adjoint of a given operator. A $\Pp$-adjoint $O^{p}$ of a given operator $O$ is given by: $O^{p} = \Pp O^{h} \Pp$. Amusingly, with this definition, one immediately sees that $\ap^{p}(q) = \am(q)$ and $\am^{p}(q) = \ap(q)$.  Thus both $N$ and $H$ become $\Pp$-Hermitean, that is, $H^{p} = H$ and $N^{p} = N$. The construction of alternative  inner products in quantised theories, in particular, involving discrete symmetries has been studied in detail; and also continues to be an area of research (for example see Ref. \cite{most1} and \cite{das}; and references therein).
Existence of Hermiticity of $N$ and $H$ is particularly welcoming since it implies that both these observables have real eigenvalues. Further, since $\psi^{h}(-x,t) = \psi^{p}(x,t)$, it turns out that: $[ \psi(x,t), \psi^{p}(y,t) ] = \delta(x-y)$. 

For brevity, in what follows, this simple notation for creation and annihilation operators is employed: $\am(p) \equiv a(p)$ and $\ap(p) \equiv a^{p}(p)$. This allows the Hamiltonian (\ref{Hamil}) to be rewritten as:
\begin{equation}
  \begin{aligned} \label{Hamil2}
  H &= H_{0} + V, \\
  H_{0} &= \int \frac{dp}{2 \pi} w_{p} a^{p}(p) a(p), \\
  V &= 2 \pi c \int \frac{d p_{1}}{2 \pi} \frac{d p_{2}}{2 \pi} \frac{d p_{3}}{2 \pi} \frac{d p_{4}}{2 \pi} \: \delta(p_{1}+p_{2}-p_{3}-p_{4}) a^{p}(p_1) a^{p}(p_2) a(p_3) a(p_4).
\end{aligned}
\end{equation}
Above expression is an important result, since it essentially means that, this theory is actually that of Bose gas with delta function interaction \cite{gutkin}. This can be seen as follows: the free Hamiltonian $H_{0}$ part simply counts the energy of nonrelativistic bosons occupying each mode. Whereas the interaction term can be written as:
\begin{equation}
  V = c \int dx dy \: \rho(x,t) \delta(x-y) \rho(y,t), 
\end{equation} 
where $\rho(x,t) = \psi^{p}(x,t) \psi(x,t)$ is particle number density, \emph{i.e.,} $N = \int dx \rho(x,t)$. This clearly shows that, the bosons interact via a pairwise delta function potential. Thus this theory is actually a theory of bosons interacting via a delta function potential with one another. Interestingly, one finds from (\ref{Hamil2}) that the equation of motion for operator $\psi$ is:
\begin{equation}
    \left( i \frac{\d}{\d t} + \frac{\d^2}{\d x^2} \right) \j = 2 c {\psi}^{p}(x,t) \j^2.
\end{equation}
This means that the quantum theory of nonlocal \nlse and that of \nlse are infact \emph{identical}. 
This result is particularly interesting, since it essentially means that this quantum theory has no nonlocal features, as the classical equation (\ref{nnlse}) might convey at the first sight.

It is well known that the quantum theory of \nlse is exactly soluble via Bethe ansatz \cite{thack, gutkin}. With this knowledge at hand, exact solution for nonlocal \nlse theory can be written easily. Infact, it is straightforward to check that, the exact two particle ``in'' eigenstate of $H$, by exactly evaluating (\ref{sol}), is given by:
\begin{equation}
  | \Psi (k_1, k_2) \rangle = - \int dx_{1} dx_{2} \: \left[ \theta(x_1 - x_2) + \theta(x_2 - x_1) e^{i \Delta(k_2 - k_1)} \right] e^{i (k_1 x_1 + k_2 x_2)} \pi(x_1) \pi(x_2) |vac \rangle, 
\end{equation}    
where $k_1 < k_2$. Here, $e^{i \Delta(k_2 - k_1)} = \frac{k_2 - k_1 - ic}{k_2 - k_1 + ic}$, and it is assumed that $c > 0$. It can be shown that, above state becomes the exact two particle ``out'' state when $k_1 > k_2$, so that the exact two particle S-matrix is $e^{i \Delta(k_2 - k_1)}$. The energy eigenvalue of both these states turn out to be $k^{2}_{1} + k^{2}_{2}$. Above result for two particle can be generalised for N-particles using the Bethe ansatz technique. Thus the exact N-particle state with energy $\sum_{i=1}^{N} k^{2}_{i}$ is given by:
\begin{align} \nonumber
  | \Phi (k_1, \cdots , k_N) \rangle = & (-i)^{N} \int \left( \prod^{N}_{i=1} e^{i k_i x_i} dx_{i} \right) \times \prod_{i < j \leq N} \left( 1 - \frac{i c}{k_i - k_j} \epsilon(x_i - x_j) \right) \\
   & \times \pi(x_1) \cdots \pi(x_N) |vac \rangle,   
\end{align}
where $\epsilon(x)$ is the signum function. Using similar approach it is possible to construct the exact N-particle states when $c < 0$, that is, when the delta function potential is attractive. In such case, it can be shown that the theory admits bound states \cite{thack}. 

\section{Conclusion}

In this paper, quantum theory of nonlocal \nlse is constructed and studied. It is observed that usual straightforward canonical route of construction leads to a theory with pathological  inner product. By redefining the inner product over the vector space, this problem is successfully overcomed. The resultant theory obtained is found to be equivalent to the quantum theory of nonrelativistic bosons interacting via delta function potential. This allows one to conclude that the quantum theory of nonlocal \nlse has no nonlocal features. Further, using the theory's equivalence to that of delta function Bose gas, the theory is shown to be exactly soluble, and exact eigenstates are found using the Bethe ansatz technique. 

A given relativistic quantum field theory of some field $\phi$, is said to be \emph{local} iff it obeys the microcausality condition: $[ \phi(x), \phi(y) ] = 0$ for all $x$ and $y$ such that $(x-y)^{2} < 0$ \cite{barton}.  This condition essentially means that the field at two points outside light cone of one another, can be accurately measured, since the effect of one measurement on the other can not travel faster than the speed of light in a local theory. Unfortunately, in the nonrelativistic quantum field theory, such a condition does not exist, due to obvious reason. The authors are of the opinion that, the locality of a theory should hence be judged by looking at the nature of interaction potential amongst the particles. This view is in agreement with the case of quantum electrodynamics in the Coulomb gauge, where the existence of instantaneous long range Coulomb potential makes the theory nonlocal \cite{mandl}.    

While the quantised theory is found to be devoid of nonlocality, the classical equation of motion for the dynamical fields remain nonlocal. Implications of this in the semiclassical regime of quantised theory requires careful examination. The quantum theory studied here can be naturally generalised to finite density and finite temperature conditions. Using similar approach, it should be possible to construct fermionic theory of nonlocal nonlinear Schr\"odinger equation as well.   

\bibliography{ref}

\begin{thebibliography}{15}%
\makeatletter
\providecommand \@ifxundefined [1]{%
 \@ifx{#1\undefined}
}%
\providecommand \@ifnum [1]{%
 \ifnum #1\expandafter \@firstoftwo
 \else \expandafter \@secondoftwo
 \fi
}%
\providecommand \@ifx [1]{%
 \ifx #1\expandafter \@firstoftwo
 \else \expandafter \@secondoftwo
 \fi
}%
\providecommand \natexlab [1]{#1}%
\providecommand \enquote  [1]{``#1''}%
\providecommand \bibnamefont  [1]{#1}%
\providecommand \bibfnamefont [1]{#1}%
\providecommand \citenamefont [1]{#1}%
\providecommand \href@noop [0]{\@secondoftwo}%
\providecommand \href [0]{\begingroup \@sanitize@url \@href}%
\providecommand \@href[1]{\@@startlink{#1}\@@href}%
\providecommand \@@href[1]{\endgroup#1\@@endlink}%
\providecommand \@sanitize@url [0]{\catcode `\\12\catcode `\$12\catcode
  `\&12\catcode `\#12\catcode `\^12\catcode `\_12\catcode `\%12\relax}%
\providecommand \@@startlink[1]{}%
\providecommand \@@endlink[0]{}%
\providecommand \url  [0]{\begingroup\@sanitize@url \@url }%
\providecommand \@url [1]{\endgroup\@href {#1}{\urlprefix }}%
\providecommand \urlprefix  [0]{URL }%
\providecommand \Eprint [0]{\href }%
\providecommand \doibase [0]{http://dx.doi.org/}%
\providecommand \selectlanguage [0]{\@gobble}%
\providecommand \bibinfo  [0]{\@secondoftwo}%
\providecommand \bibfield  [0]{\@secondoftwo}%
\providecommand \translation [1]{[#1]}%
\providecommand \BibitemOpen [0]{}%
\providecommand \bibitemStop [0]{}%
\providecommand \bibitemNoStop [0]{.\EOS\space}%
\providecommand \EOS [0]{\spacefactor3000\relax}%
\providecommand \BibitemShut  [1]{\csname bibitem#1\endcsname}%
\let\auto@bib@innerbib\@empty
\bibitem [{\citenamefont {Ablowitz}\ and\ \citenamefont
  {Musslimani}(2013)}]{ablo}%
  \BibitemOpen
  \bibfield  {author} {\bibinfo {author} {\bibfnamefont {M.~J.}\ \bibnamefont
  {Ablowitz}}\ and\ \bibinfo {author} {\bibfnamefont {Z.~H.}\ \bibnamefont
  {Musslimani}},\ }\href@noop {} {\bibfield  {journal} {\bibinfo  {journal}
  {Physical Review Letters}\ }\textbf {\bibinfo {volume} {110}},\ \bibinfo
  {pages} {064105} (\bibinfo {year} {2013})}\BibitemShut {NoStop}%
\bibitem [{\citenamefont {Bogoliubov}\ and\ \citenamefont
  {Shirkov}(1959)}]{bogol}%
  \BibitemOpen
  \bibfield  {author} {\bibinfo {author} {\bibfnamefont {N.~N.}\ \bibnamefont
  {Bogoliubov}}\ and\ \bibinfo {author} {\bibfnamefont {D.~V.}\ \bibnamefont
  {Shirkov}},\ }\href@noop {} {\emph {\bibinfo {title} {Introduction to the
  theory of quantized fields}}},\ Vol.~\bibinfo {volume} {59}\ (\bibinfo
  {publisher} {Interscience New York},\ \bibinfo {year} {1959})\BibitemShut
  {NoStop}%
\bibitem [{Note1()}]{Note1}%
  \BibitemOpen
  \bibinfo {note} {In this manuscript throughout, subscript stands for partial
  differentiation with respect to the variable, for example $f_{x}(x,t) =
  \protect \frac {\partial f(x,t)}{\partial x}$.}\BibitemShut {Stop}%
\bibitem [{\citenamefont {Schwinger}(1951)}]{schwing}%
  \BibitemOpen
  \bibfield  {author} {\bibinfo {author} {\bibfnamefont {J.}~\bibnamefont
  {Schwinger}},\ }\href@noop {} {\bibfield  {journal} {\bibinfo  {journal}
  {Physical Review}\ }\textbf {\bibinfo {volume} {82}},\ \bibinfo {pages} {914}
  (\bibinfo {year} {1951})}\BibitemShut {NoStop}%
\bibitem [{\citenamefont {Abhinav}\ \emph {et~al.}(2013)\citenamefont
  {Abhinav}, \citenamefont {Jayannavar},\ and\ \citenamefont
  {Panigrahi}}]{abhi}%
  \BibitemOpen
  \bibfield  {author} {\bibinfo {author} {\bibfnamefont {K.}~\bibnamefont
  {Abhinav}}, \bibinfo {author} {\bibfnamefont {A.}~\bibnamefont {Jayannavar}},
  \ and\ \bibinfo {author} {\bibfnamefont {P.~K.}\ \bibnamefont {Panigrahi}},\
  }\href@noop {} {\bibfield  {journal} {\bibinfo  {journal} {Annals of
  Physics}\ }\textbf {\bibinfo {volume} {331}},\ \bibinfo {pages} {110}
  (\bibinfo {year} {2013})}\BibitemShut {NoStop}%
\bibitem [{\citenamefont {Dirac}(1967)}]{dir}%
  \BibitemOpen
  \bibfield  {author} {\bibinfo {author} {\bibfnamefont {P.~A.~M.}\
  \bibnamefont {Dirac}},\ }\href@noop {} {\emph {\bibinfo {title} {The
  principles of Quantum Mechanics}}}\ (\bibinfo  {publisher} {Oxford University
  Press},\ \bibinfo {year} {1967})\BibitemShut {NoStop}%
\bibitem [{\citenamefont {Sakurai}(1994)}]{sakurai}%
  \BibitemOpen
  \bibfield  {author} {\bibinfo {author} {\bibfnamefont {J.~J.}\ \bibnamefont
  {Sakurai}},\ }\href@noop {} {\emph {\bibinfo {title} {Modern Quantum
  Mechanics}}}\ (\bibinfo  {publisher} {Addison-Wesley Pub. Co},\ \bibinfo
  {year} {1994})\BibitemShut {NoStop}%
\bibitem [{\citenamefont {Heisenberg}(1957)}]{hei}%
  \BibitemOpen
  \bibfield  {author} {\bibinfo {author} {\bibfnamefont {W.}~\bibnamefont
  {Heisenberg}},\ }\href@noop {} {\bibfield  {journal} {\bibinfo  {journal}
  {Nuclear Physics}\ }\textbf {\bibinfo {volume} {4}},\ \bibinfo {pages} {532}
  (\bibinfo {year} {1957})}\BibitemShut {NoStop}%
\bibitem [{\citenamefont {Barton}(1963)}]{barton}%
  \BibitemOpen
  \bibfield  {author} {\bibinfo {author} {\bibfnamefont {G.}~\bibnamefont
  {Barton}},\ }\href@noop {} {\emph {\bibinfo {title} {Introduction to advanced
  field theory}}}\ (\bibinfo  {publisher} {Wiley},\ \bibinfo {year}
  {1963})\BibitemShut {NoStop}%
\bibitem [{\citenamefont {Lee}\ and\ \citenamefont {Wick}(1969)}]{lee}%
  \BibitemOpen
  \bibfield  {author} {\bibinfo {author} {\bibfnamefont {T.}~\bibnamefont
  {Lee}}\ and\ \bibinfo {author} {\bibfnamefont {G.}~\bibnamefont {Wick}},\
  }\href@noop {} {\bibfield  {journal} {\bibinfo  {journal} {Nuclear Physics
  B}\ }\textbf {\bibinfo {volume} {9}},\ \bibinfo {pages} {209} (\bibinfo
  {year} {1969})}\BibitemShut {NoStop}%
\bibitem [{\citenamefont {Mostafazadeh}(2003)}]{most1}%
  \BibitemOpen
  \bibfield  {author} {\bibinfo {author} {\bibfnamefont {A.}~\bibnamefont
  {Mostafazadeh}},\ }\href@noop {} {\bibfield  {journal} {\bibinfo  {journal}
  {Journal of Physics A: Mathematical and General}\ }\textbf {\bibinfo {volume}
  {36}},\ \bibinfo {pages} {7081} (\bibinfo {year} {2003})}\BibitemShut
  {NoStop}%
\bibitem [{\citenamefont {Das}\ and\ \citenamefont {Greenwood}(2009)}]{das}%
  \BibitemOpen
  \bibfield  {author} {\bibinfo {author} {\bibfnamefont {A.}~\bibnamefont
  {Das}}\ and\ \bibinfo {author} {\bibfnamefont {L.}~\bibnamefont
  {Greenwood}},\ }\href@noop {} {\bibfield  {journal} {\bibinfo  {journal}
  {Physics Letters B}\ }\textbf {\bibinfo {volume} {678}},\ \bibinfo {pages}
  {504} (\bibinfo {year} {2009})}\BibitemShut {NoStop}%
\bibitem [{\citenamefont {Gutkin}(1988)}]{gutkin}%
  \BibitemOpen
  \bibfield  {author} {\bibinfo {author} {\bibfnamefont {E.}~\bibnamefont
  {Gutkin}},\ }\href@noop {} {\bibfield  {journal} {\bibinfo  {journal}
  {Physics Reports}\ }\textbf {\bibinfo {volume} {167}},\ \bibinfo {pages} {1}
  (\bibinfo {year} {1988})}\BibitemShut {NoStop}%
\bibitem [{\citenamefont {Thacker}(1981)}]{thack}%
  \BibitemOpen
  \bibfield  {author} {\bibinfo {author} {\bibfnamefont {H.~B.}\ \bibnamefont
  {Thacker}},\ }\href@noop {} {\bibfield  {journal} {\bibinfo  {journal}
  {Reviews of Modern Physics}\ }\textbf {\bibinfo {volume} {53}},\ \bibinfo
  {pages} {253} (\bibinfo {year} {1981})}\BibitemShut {NoStop}%
\bibitem [{\citenamefont {Mandl}\ and\ \citenamefont {Shaw}(2010)}]{mandl}%
  \BibitemOpen
  \bibfield  {author} {\bibinfo {author} {\bibfnamefont {F.}~\bibnamefont
  {Mandl}}\ and\ \bibinfo {author} {\bibfnamefont {G.}~\bibnamefont {Shaw}},\
  }\href@noop {} {\emph {\bibinfo {title} {Quantum field theory}}}\ (\bibinfo
  {publisher} {John Wiley \& Sons},\ \bibinfo {year} {2010})\BibitemShut
  {NoStop}%
\end{thebibliography}%

\end{document}